\documentclass[11pt]{article}
\usepackage{amssymb,latexsym,color,amsmath,pifont,epsfig,cite}
\usepackage{setspace,mathtools}
     \mathtoolsset{showmanualtags}

\newtheorem{theorem}{Theorem}
\newtheorem{corollary}[theorem]{Corollary}
\newtheorem{lemma}[theorem]{Lemma}

\newtheorem{example}{Example}

\newcommand{\enma}[1]   {\ensuremath{#1}}

\newcommand{\beq}{\begin{equation}}
\newcommand{\eeq}{\end{equation}}
\newcommand{\beqn}{\begin{eqnarray}}
\newcommand{\eeqn}{\end{eqnarray}}
\newcommand{\bseq}{\begin{subequations}}
\newcommand{\eseq}{\end{subequations}}
\newcommand{\ba}{\begin{array}}
\newcommand{\ea}{\end{array}}
\newcommand{\bct}{\begin{center}}
\newcommand{\ect}{\end{center}}
\newcommand{\btmz}{\begin{itemize}}
\newcommand{\etmz}{\end{itemize}}
\newcommand{\benum}{\begin{enumerate}}
\newcommand{\eenum}{\end{enumerate}}

\newcommand{\R}{{\mathbb R}}

\newcommand{\norm}[1]{\| #1 \|}                 

\newcommand{\sgn}       {\enma{\mathrm{sgn}}}

\newcommand{\be}{\begin{equation}}
\newcommand{\ee}{\end{equation}}

\newcommand{\cplxs}{ C\kern -.35em \rule{0.03 em}{.7 ex}~   }

\def\complex{\hbox{C\kern -.45em \rule{0.03 em}{1.5 ex}}~}

\newcommand{\bi}{\begin{itemize}}
\newcommand{\ei}{\end{itemize}}

\def \RR {I \! \! R}

\newcommand{\bbR}{\mathbb{R}}

\newcommand{\ip}[2]{\langle #1 , #2 \rangle}
\newcommand{\ipT}[2]{\langle #1, #2 \rangle_T}

\newcommand{\normT}[1]{\left\|#1\right\|_T}

\begin{document}
\title{A Passivity-Based Stability Criterion for a Class of Interconnected Systems and
 Applications to Biochemical Reaction Networks}
\author{\begin{minipage}[t]{0.5\textwidth}\begin{center} {\Large Murat
Arcak} \\  Department of Electrical, Computer, and Systems Engineering\\
Rensselaer Polytechnic Institute, Troy, NY \\ {\tt arcakm@rpi.edu}
\end{center}\end{minipage}\begin{minipage}[t]{0.5\textwidth}\begin{center}
{\Large Eduardo D. Sontag}
\\  Department of Mathematics \\ Rutgers
University, New Brunswick, NJ \\ {\tt sontag@math.rutgers.edu}
\end{center}\end{minipage} }
\maketitle

\thispagestyle{empty}

\begin{abstract}This paper presents a stability test for a class
of interconnected nonlinear systems motivated by biochemical
reaction networks. The main result  determines global asymptotic
stability of the network from the diagonal stability of a {\it
dissipativity matrix} which incorporates information about the
passivity properties of the subsystems, the interconnection
structure of the network, and the signs of the interconnection
terms. This stability test encompasses the {\it secant criterion}
for cyclic networks presented in \cite{arcson06}, and extends it to
a general interconnection structure represented by a graph. The new
stability test is illustrated on a mitogen activated protein kinase
(MAPK) cascade model, and on a branched interconnection structure
motivated by metabolic networks. The next problem addressed is the
robustness of stability in the presence of diffusion terms. The
authors use a compartmental model to represent the localization of
the reactions and present conditions under which stability is
preserved despite the diffusion terms between the compartments.
\end{abstract}

\section{Introduction}
This paper continues the development of passivity-based stability
criteria for interconnected systems motivated by classes of
biochemical reaction networks. In \cite{secant,arcson06} the authors
studied a cyclic interconnection structure in which the first
subsystem of a cascade is driven by a negative feedback  from the
last subsystem downstream.  This cyclic feedback structure is
ubiquitous  in gene regulation networks \cite{Goodwin65,
TiwariFraserBeckmann74, FraserTiwari74, HasTysWeb77,
SanglierNicolis76, tysoth78, BanksMahaffy781, BanksMahaffy782,
Mahaffy84, Mahaffy842, Glass-oscill, thr91}, cellular signaling
pathways \cite{KholodenkoNegfeedback,stas2}, and has also been noted
in metabolic pathways \cite{MorKay67,stephanopoulous}. In
\cite{secant,arcson06}  the authors  first presented a passivity
interpretation of the ``secant criterion" developed earlier  in
\cite{tysoth78,thr91} for the stability of linear cyclic systems,
and next used this passivity insight to extend the secant criterion
to nonlinear systems. The notion of {\it passivity} evolved from an
abstraction of energy conservation and dissipation in electrical and
mechanical systems \cite{Wil72,vds}, into a fundamental tool
routinely used for nonlinear system design and analysis
\cite{sepulchre,AK01survey}.

The first contribution of this paper is to expand the analysis tool
of \cite{arcson06} to a general interconnection structure, thus
obtaining a broadly applicable stability criterion that encompasses
the secant criterion for cyclic systems as a special case. As in
\cite{arcson06}, our approach is to exploit the {passivity}
properties  and the corresponding {\it storage functions}
\cite{Wil72} for smaller components that comprise the network, and
to construct a composite Lyapunov function for the interconnection
using these storage functions. The idea of using composite Lyapunov
functions has been explored extensively in the literature of
large-scale systems as surveyed in \cite{michel, siljak}, and led to
several network small-gain criteria \cite{araki75,araki76} that
restrict the strength of the interconnection terms. A distinguishing
feature of our passivity-based criterion, however, is that we take
advantage of the sign properties of the interconnection terms to
obtain less conservative stability conditions than the small-gain
approach.

To determine the stability of the resulting network of passive
subsystems we follow the formalism of \cite{MoyHill78,vidyasagar2},
and construct a {\it dissipativity matrix} (denoted by $E$ below)
that incorporates information about the passivity properties of the
subsystems, the interconnection structure of the network, and the
signs of the interconnection terms. As a stability test for the
interconnected system, we check the {\it diagonal stability}
\cite{bhaya} of this dissipativity matrix, that is, the existence of
a diagonal solution $D>0$ to the Lyapunov equation $E^TD+DE<0$
which, if feasible, proves that the network is indeed stable. In
particular, the diagonal entries of $D$ serve as the weights of the
storage functions in our composite Lyapunov function. Although
similar results can be proven by combining the pure input/output
approach in \cite{MoyHill78,vidyasagar2} with appropriate
detectability and controllability conditions (see the discussion in
Section \ref{MoyHillVid} below),
 the direct Lyapunov
approach employed in this paper allows us to formulate verifiable
state-space conditions that guarantee the desired passivity
properties for the subsystems. These conditions are particularly
suitable for systems of biological interest because they are
applicable to models with nonnegative state variables, and do not
rely on the knowledge of the location of the equilibrium.

The second contribution of this paper is to accommodate state
products which are disallowed in the nonlinear model studied in
\cite{arcson06}. This is achieved with a new storage function
construction for each subsystem which, in the absence of state
products, coincides with the construction in \cite{arcson06}. Thanks
to this extension, our stability criterion is now applicable to a
broader class of models, even in the case of cyclic systems. This
class encompasses a mitogen activated protein kinase (MAPK) cascade
model with inhibitory feedback proposed in
\cite{KholodenkoNegfeedback,stas2}, which is studied in Example
\ref{mapk} as an illustration of our main result. The final result
in the paper employs a compartmental model to describe the spatial
localization of the reactions, and proves that, if the
passivity-based stability criterion holds for each compartment and
if the storage functions satisfy an additional convexity property,
then stability is preserved in the presence of diffusion terms
between the compartments.

The paper is organized as follows: Section \ref{recap} gives an
overview of the main results in \cite{arcson06}. Section
\ref{mainsec} presents a general interconnection structure
represented by a graph, and gives the main stability result of the
paper. Section \ref{examp} illustrates this result on biologically
motivated examples. Section \ref{compsec} studies robustness of
stability in the presence of diffusion terms in a compartmental
model. Section \ref{MoyHillVid} develops an extension of the purely
input/output-based proof in \cite{secant} of the secant criterion to
the general graphs studied in this paper. In doing so, it adapts a
lemma from \cite{MoyHill78,vidyasagar2} and compares the
input/output approach with the state-space results derived earlier
in the paper. Section \ref{conc} gives the conclusions.

\section{Overview of the Secant Criterion for Cyclic
Systems}\label{recap} To evaluate stability properties of negative
feedback cyclic systems, references \cite{tysoth78,thr91} analyzed
the Jacobian linearization at the equilibrium, which is of the form
\begin{equation}
    A \; = \;
    \left[\begin{array}{ccccc} -a_1  & 0 & \cdots & 0 & -b_n \\ b_1 & -a_2 &\ddots & & 0 \\
    0 & b_2 & -a_3 & \ddots & \vdots \\ \vdots & \ddots & \ddots &
    \ddots & 0 \\ 0 & \cdots & 0 & b_{n-1} & -a_n
    \end{array} \right]
    \label{Amatrix}
    \end{equation}
$a_i>0,
    ~
    b_i>0,
    ~
    i=1,\cdots,n$, and showed that $A$ is Hurwitz if the following
    sufficient condition
holds:
    \begin{equation}
    \label{eq.secant}
    \frac{b_1\cdots b_n}{a_1\cdots a_n} \, < \, \sec(\pi/n)^n.
    \end{equation}
Unlike a {\it small-gain} condition which would restrict the
right-hand side of (\ref{eq.secant}) to be  $1$, the ``secant
criterion'' (\ref{eq.secant}) also exploits the phase of the loop
and allows the right-hand side to be as high as $8$ (when $n=3$).
The secant criterion  is also necessary for stability when the
$a_i$'s are identical.

\begin{figure}[ht]
\centering
\includegraphics[scale=0.35]{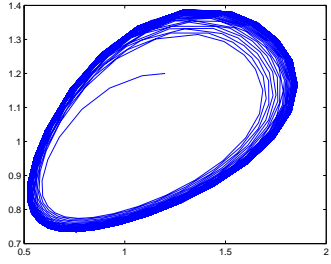}
\caption{Trajectory of (\ref{counterex}) starting from initial
condition $x=[1.2\ 1.2\ 1.2]^T$, projected onto the $x_1$-$x_2$
plane.} \label{USE}
\end{figure}

Local stability of the equilibrium proven in \cite{tysoth78,thr91},
however, does not rule out the possibility of periodic orbits.
Indeed, the Poincar\'{e}-Bendixson Theorem of Mallet-Paret and Smith
 for cyclic systems \cite{malletparet-smith, Mallet-ParetDelay}
allows such periodic orbits to coexist with stable equilibria, as we
illustrate on the system:
\begin{eqnarray}\nonumber
\dot{x}_1&=&-x_1+\varphi(x_3)\\
\dot{x}_2&=&-x_2+x_1 \label{counterex} \\
\dot{x}_3&=&-x_3+x_2 \nonumber
\end{eqnarray}
where
\begin{equation}\label{sat}
\varphi(x_3)=e^{-10(x_3-1)}+0.1{\rm sat}(25(x_3-1)),
\end{equation}
and ${\rm sat}(\cdot):=\sgn(\cdot)\min\{1,|\cdot|\}$ is a
saturation\footnote{One can easily modify this example to make
$\varphi(\cdot)$ smooth while retaining the same stability
properties.} function. The function (\ref{sat}) is decreasing, and
its slope has magnitude $b_3=7.5$ at the equilibrium
$x_1=x_2=x_3=1$. With $a_1=a_2=a_3=b_1=b_2=1$ and $n=3$, the secant
criterion (\ref{eq.secant}) is satisfied and, thus, the equilibrium
is asymptotically stable. However, simulations in Figure \ref{USE}
show the existence of a periodic orbit in addition to this stable
equilibrium. 

To study  {\em global} stability properties of cyclic systems with
negative feedback, in \cite{secant,arcson06} the authors first
developed a passivity interpretation of the secant criterion
(\ref{eq.secant}), and next used this passivity insight to extend
the secant criterion to the nonlinear model:
    \begin{eqnarray}\nonumber
    \dot{x}_1
    & \!\! = \!\! &
    -f_1(x_1) \, + \, h_n(x_n)
    \\ \nonumber
    \dot{x}_2
    & \!\! = \!\! &
    -f_2(x_2) \, + \, h_1(x_1)
    \\ \label{bio}
    &\vdots&
    \\ \nonumber
    \dot{x}_n
    & \!\! = \!\! &
    -f_n(x_n) \, + \, h_{n-1}(x_{n-1})
    \end{eqnarray}
in which $x_i\in \RR_{\ge 0}$, $f_i(\cdot)$, $i=1,\cdots,n$ and
$h_i(\cdot)$, $i=1,\cdots,n-1$ are increasing functions, and
$h_n(\cdot)$ is a decreasing function which represents the
inhibition of the formation of $x_1$ by the end product $x_n$. When
an equilibrium $x^*$ exists, \cite{arcson06} proves its global
asymptotic stability under the following condition:
\begin{eqnarray}
&&\frac{\left|\frac{\partial h_i(x_i)}{\partial
x_i}\right|}{\frac{\partial f_i(x_i)}{\partial x_i}}\le \gamma_i
\quad \forall x_i \in \RR_{\ge 0}, \quad i=1,\cdots,n, \label{inc2b}\\
&& \gamma_1\cdots\gamma_n \, < \, \sec(\pi/n)^n,\label{inc2c}
\end{eqnarray}
which encompasses the linear secant criterion (\refeq{eq.secant})
with $\gamma_i=b_i/a_i$.

A crucial ingredient in the global asymptotic stability
 proof of \cite{arcson06} is the observation that the secant
 condition (\ref{inc2c}) is  necessary and sufficient  for
 the
 {\em diagonal stability\/} of the matrix
\begin{equation}
    E_{cyclic} \; = \;
    \left[\begin{array}{ccccc} -1/\gamma_1  & 0 & \cdots & 0 & -1 \\ 1 & -1/\gamma_2 &\ddots & & 0 \\
    0 & 1 & -1/\gamma_3 & \ddots & \vdots \\ \vdots & \ddots & \ddots &
    \ddots & 0 \\ 0 & \cdots & 0 & 1 & -1/\gamma_n
    \end{array} \right]
    \label{Ematrix}
    \end{equation}
  that is, for the existence of
 a diagonal matrix $D>0$ such that
 \begin{equation}\label{cycdiag}
 E_{cyclic}^T D \; + \; D E_{cyclic}\; < \; 0.
 \end{equation}
The authors of \cite{arcson06} connect this diagonal stability
property to the global asymptotic stability of (\ref{bio}) by first
noting that the assumption (\ref{inc2b}) guarantees an {\it output
strict passivity} property, where $1/\gamma_i$ quantifies the excess
of passivity in each $x_i$-subsystem. They then incorporate this
passivity information in the diagonal terms of the
\textit{dissipativity matrix} (\ref{Ematrix}), and represents the
interconnection structure with the off-diagonal terms. Finally, they
use the diagonal stability condition (\ref{cycdiag}), which is
equivalent to the secant criterion (\ref{inc2c}), to check whether
the excess of passivity in each subsystem overcomes the loss of
passivity in the interconnection. In particular, the  diagonal
entries of $D$ constitute the weights of the storage functions in a
composite Lyapunov function for (\ref{bio}).

\section{From the Cyclic Structure to General Graphs}\label{mainsec}
We now extend the diagonal stability procedure outlined above for
cyclic systems to a general interconnection structure, described by
a directed graph without self-loops.  The nodes represent subsystems
with possibly vector outputs, and a separate link is used for each
output channel. For the nodes $i=1,\cdots,N$ and links
$l=1,\cdots,M$,
 we denote by $\mathcal{L}^+_i\subseteq \{1,\cdots,M\}$ the subset of links
 for which node $i$ is the sink, and by
$\mathcal{L}^-_i$ the subset of links for which node $i$ is the
source. We write $i={\rm source}(l)$ if $l\in \mathcal{L}^-_{i}$,
and $i={\rm sink}(l)$ if $l\in \mathcal{L}^+_{i}$. Using this graph
we introduce the dynamic system:
\begin{equation}\label{newsys}
\dot{x}_i=-f_i(x_i)+g_i(x_i)\sum_{l\in \mathcal{L}^+_i}h_l(x_{{\rm
source}(l)}) \quad i=1,\cdots,N
\end{equation}
where $x_i\in \RR_{\ge 0}$, and $f_i(\cdot)$, $g_i(\cdot)$,
$i=1,\cdots,N$, $h_l(\cdot)$, $l=1,\cdots,M$ are locally Lipschitz
functions further restricted by the following assumptions:

{\it A1:} $f_i(0)=0$ and, for all $\sigma\ge 0$, $g_i(\sigma)>0$,
$h_l(\sigma)\ge 0$.

Assumption A1 guarantees that the nonnegative orthant $\RR^N_{\ge
0}$ is forward invariant for (\ref{newsys}). The strict positivity
of $g_i(x_i)$ is also essential for our analysis since we exploit
the sign properties of $h_l(x_{{\rm source}(l)})$ which are
multiplied by $g_i(x_i)$ in (\ref{newsys}).

{\it A2:} There exists an equilibrium $x^*\in \RR^N_{\ge 0}$ for
(\ref{newsys}).

{\it A3:} For each node $i$, the function $f_i(x_i)/g_i(x_i)$
satisfies the {\it sector} property:
\begin{equation}\label{a3}
(x_i-x_i^*)\left(\frac{f_i(x_i)}{g_i(x_i)}-\frac{f_i(x^*_i)}{g_i(x^*_i)}\right)
> 0\quad \forall x_i \in \RR_{\ge 0}-\{x_i^*\}.
\end{equation}

{\it A4:} For each node $i$, and for each link $l\in
\mathcal{L}_i^-$, the function $h_l(x_i)$ satisfies one of the
following {sector} properties for all $x_i \in \RR_{\ge
0}-\{x_i^*\}$:
\begin{eqnarray}\label{pos}
&& (x_i-x_i^*)[h_l(x_i)-h_l(x_i^*)]> 0 \\
&&(x_i-x_i^*)[h_l(x_i)-h_l(x_i^*)]< 0. \label{neg}
\end{eqnarray}

To distinguish between positive and negative feedback signals we
assign to each link $l$ a positive sign if (\ref{pos}) holds, and a
negative sign if (\ref{neg}) holds, and rewrite
(\ref{pos})-(\ref{neg}) as
\begin{equation}\label{compact}
{\rm sign}({\rm link }\ l)(x_i-x_i^*)[h_l(x_i)-h_l(x_i^*)]> 0
\end{equation}
$\forall x_i \in \RR_{\ge 0}-\{x_i^*\}$.

{\it A5:} For each link $l\in \mathcal{L}_i^-$ there exists a
constant $\gamma_l > 0$ such that, $\forall x_i \in \RR_{\ge
0}-\{x_i^*\}$,
\begin{equation}\label{gamma}
{\rm sign}({\rm link }\
l)\frac{h_l(x_i)-h_l(x_i^*)}{\frac{f_i(x_i)}{g_i(x_i)}-\frac{f_i(x^*_i)}{g_i(x^*_i)}}\le
\gamma_l.
\end{equation}

\begin{theorem}\label{main}\rm
 Consider the system (\ref{newsys}), and suppose assumptions A1-A5 hold. If the $M\times M$ dissipativity matrix
\begin{equation}\label{newE}
E_{lk}=\left\{\begin{array}{ll} -1/\gamma_l & \mbox{if} $\ k=l$
\\{\rm sign}({\rm link}\ k) & \mbox{if source$(l)=\
$sink$(k)$} \\ 0 & {\rm otherwise}
\end{array} \right.
\end{equation}
is diagonally stable; that is, if there exists a diagonal matrix
$D>0$ such that
\begin{equation}\label{diagstab}
 E^T D \; + \; D E\; < \; 0,
 \end{equation}
 then the equilibrium $x^*$ is asymptotically stable. If, further,
 for each node $i$ one of the following two conditions holds, then $x^*$ is globally asymptotically stable in
$\RR^N_{\ge 0}$:

a) $\mathcal{L}_i^-$ is nonempty and
 there exists at least one link $l\in
 \mathcal{L}_i^-$ such that
 \begin{equation}\label{integ}
\lim_{x_i\rightarrow
\infty}\int_{x_i^*}^{x_i}\frac{h_l(\sigma)-h_l(x_i^*)}{g_i(\sigma)}
=\infty,
\end{equation}

b) $\mathcal{L}_i^-$ is empty; that is, the outdegree of node $i$ is
zero;
\begin{equation}\label{integ2}
\lim_{x_i\rightarrow
\infty}\int_{x_i^*}^{x_i}\frac{\sigma-x_i^*}{g_i(\sigma)} =\infty,
\end{equation}
and there exists a class-$\mathcal{K}_\infty$
function\footnote{$\mathcal{K}$ is the class of
  functions $\RR_{\ge 0}\rightarrow \RR_{\ge 0}$ which are zero at zero,
  strictly increasing and continuous. $\mathcal{K}_\infty$ is the subset of
  class-$\mathcal{K}$ functions that are unbounded.} $\omega(\cdot)$ such that
\begin{equation}\label{a3s}
(x_i-x_i^*)\left(\frac{f_i(x_i)}{g_i(x_i)}-\frac{f_i(x^*_i)}{g_i(x^*_i)}\right)
\ge |x_i-x^*_i|\,\omega(|x_i-x^*_i|) \quad \forall x_i\ge 0.
\end{equation}
\end{theorem}

\noindent \emph{Proof:}  We first prove the theorem for the case
when $\mathcal{L}_i^-$ is nonempty for all $i=1,\cdots,N$; that is,
when there are no nodes with outdegree equal to zero. In this case
we construct a composite Lyapunov function of the form
\begin{equation}\label{lyap}
V(x-x^*)=\sum_{l=1}^M d_l V_l(x_{{\rm source}(l)}-x_{{\rm
source}(l)}^*)
\end{equation}
in which the components are
\begin{equation}\label{comp}
V_l(x_{{\rm source}(l)}-x_{{\rm source}(l)}^*)= {\rm sign}({\rm link
}\ l)\int_{x^*_{{\rm source}(l)}}^{x_{{\rm
source}(l)}}\frac{h_l(\sigma)-h_l(x^*_{{\rm source}(l)})}{g_{{\rm
source}(l)}(\sigma)}d\sigma
\end{equation}
and
 the coefficients $d_l>0$ are to be
determined. The function (\ref{lyap}) is positive definite because
each component $V_l$ is a positive definite function of $(x_{{\rm
source}(l)}-x_{{\rm source}(l)}^*)$
 due to the sign property (\ref{compact})
of the integrand in (\ref{comp}), and because  $(x_{{\rm
source}(l)}-x_{{\rm source}(l)}^*)=0$, $l=1,\cdots,M$, guarantees
$x-x^*=0$ by virtue of the fact that each node is the source for at
least one link.

We now claim that the function $V_l$ in (\ref{comp}) satisfies the
dissipativity property
\begin{equation}\label{est}
\dot{V}_l\le y_l\sum_{k=1}^M E_{lk}y_k
\end{equation}
where
\begin{equation}\label{ydef}
{y}_l:={\rm sign}({\rm link }\ l)[h_l(x_{{\rm
source}(l)})-h_l(x_{{\rm source}(l)}^*)]
\end{equation}
$l=1,\cdots,M$, and the coefficients $E_{lk}$ are as in
(\ref{newE}). Before we prove this claim, we first note that the
diagonal stability property (\ref{diagstab}) and the estimate
(\ref{est}) together imply that the Lyapunov function (\ref{lyap}),
with coefficients $d_l$ obtained from the diagonal elements of $D$,
yields a negative definite derivative from which asymptotic
stability of $x^*$ follows. If, further,
 for each node $i$ there exists at least one link $l\in
 \mathcal{L}_i^-$ such that (\ref{integ})
holds, then the Lyapunov function (\ref{lyap}) grows unbounded as
$|x|\rightarrow \infty$, thus proving global asymptotic stability.

If there exist nodes with outdegree equal to zero, then the
arguments above prove that the subsystem comprising of the nodes
with outdegree  one or more is asymptotically stable. The outputs
$h_l$ from this subsystem serve as inputs to the nodes with
outdegree equal to zero. Because the dynamics of these nodes are of
the form (\ref{newsys}) and are asymptotically stable by A3,
asymptotic stability for the equilibrium $x^*$  follows from
standard results on  cascade interconnections of asymptotically
stable systems (see {\it e.g.} \cite[p. 275]{hahn}). Likewise, when
condition (b) holds, (\ref{a3s}) and (\ref{integ2}) imply an {\it
input-to-state stability} (ISS) property \cite{S89a} for the driven
subsystem of the cascade, and global asymptotic stability follows
because the cascade interconnection of an ISS system driven by a
globally asymptotically stable system is globally asymptotically
stable \cite{S89a}.

We conclude the proof by showing that the claim (\ref{est}) is
indeed true. To this end we compute from (\ref{comp}) and
(\ref{newsys}) the derivative
\begin{equation}\label{derivative}
\dot{V}_l={\rm sign}({\rm link }\
l)[h_l(x_i)-h_l(x_i^*)]\left(-\frac{f_i(x_i)}{g_i(x_i)}+u_i\right)
\end{equation}
where $i={\rm source}(l)$, and
\begin{equation}\label{udef}
u_i:=\sum_{k\in \mathcal{L}^+_i}h_k(x_{{\rm source}(k)}).
\end{equation}
Adding and subtracting
\begin{equation}\label{ustardef}
u^*_i:=\sum_{k\in \mathcal{L}^+_i}h_k(x^*_{{\rm
source}(k)})=\frac{f_i(x^*_i)}{g_i(x^*_i)}
\end{equation}
within the bracketed term in (\ref{derivative}), we obtain
\begin{equation}
\dot{V}_l={\rm sign}({\rm link }\
l)[h_l(x_i)-h_l(x_i^*)]\left(-\frac{f_i(x_i)}{g_i(x_i)}+\frac{f_i(x^*_i)}{g_i(x^*_i)}+u_i-u_i^*\right).\label{derivative2}
\end{equation}
Next, noting that ${\rm sign}({\rm link }\ l)[h_l(x_i)-h_l(x_i^*)]$
and
$\left(\frac{f_i(x_i)}{g_i(x_i)}-\frac{f_i(x^*_i)}{g_i(x^*_i)}\right)$
possess the same signs due to (\ref{a3}) and (\ref{compact}), and
using (\ref{gamma}), we obtain the inequality
\begin{equation} -{\rm sign}({\rm link }\ l)[h_l(x_i)-h_l(x_i^*)]\left(\frac{f_i(x_i)}{g_i(x_i)}-\frac{f_i(x^*_i)}{g_i(x^*_i)}\right)\le
-\frac{1}{\gamma_l}[h_l(x_i)-h_l(x_i^*)]^2.\label{ineq}
\end{equation}
Substituting (\ref{ineq}) in (\ref{derivative2}), and using the
variables $y_l$ defined in (\ref{ydef}), we get
\begin{equation}\label{derivative3}
\dot{V}_l=-\frac{1}{\gamma_l}y_l^2+y_l(u_i-u_i^*).
\end{equation}
Finally, noting from (\ref{udef}) and (\ref{ustardef}) that
\begin{equation}\label{input}
u_i-u_i^*=\sum_{k\in \mathcal{L}_i^+}{\rm sign}({\rm link }\ k)y_k,
\end{equation}
we rewrite (\ref{derivative3}) as
\begin{equation}
\dot{V}_l\le -\frac{1}{\gamma_l}y_l^2+y_l\sum_{k\in
\mathcal{L}_i^+}{\rm sign}({\rm link }\ k)y_k,
\end{equation}
which is equivalent to (\ref{est}) by the definition of the
coefficients $E_{kl}$ in (\ref{newE}). \hfill $\Box$

\medskip

The assumptions A3-A5 rely on the knowledge of the equilibrium $x^*$
which may not be available in practice. When the functions
$f_i(\cdot)$, $g_i(\cdot)$, and $h_l(\cdot)$ are $C^1$, the
following incremental conditions guarantee A3-A5, and do not depend
on $x^*$:

{\it A3':} For each $i=1,\cdots,N$, and $ \forall x_i\ge 0$,
\begin{equation}
\frac{\partial}{\partial
x_i}\left(\frac{f_i(x_i)}{g_i(x_i)}\right)>0.
\end{equation}

{\it A4':} For each $l=1,\cdots,M$, and $ \forall x_i \ge 0$,
\begin{equation}\label{recp}
{\rm sign}({\rm link} \ l)\frac{\partial h_l(x_i)}{\partial x_i}>0.
\end{equation}

{\it A5':} For each link $l\in \mathcal{L}_i^-$ there exists a
constant $\gamma_l > 0$ such that
\begin{equation}\label{growth}
\frac{\left|\frac{\partial h_l(x_i)}{\partial x_i}
\right|}{\frac{\partial}{\partial
x_i}\left(\frac{f_i(x_i)}{g_i(x_i)}\right)}  \le \gamma_l
   \quad \forall x_i\ge 0.
\end{equation}

Although the growth assumption (\ref{growth}) may appear
restrictive, most biologically relevant nonlinearities satisfy this
condition globally. If there exist closed intervals $\mathcal{X}_i
\subseteq \RR_{\ge 0}$ such that $\mathcal{X}_1\times \cdots \times
\mathcal{X}_N$ is forward invariant for (\ref{newsys}), a less
conservative $\gamma_l$ may be obtained by evaluating (\ref{growth})
on $\mathcal{X}_i$, rather than for all $x_i\ge 0$. This relaxation
is particularly useful in biological applications where $x_i$
represents the amount of a substance which may be lower- and
upper-bounded.

The dissipativity matrix $E$ in (\ref{newE}) combines information
about the interconnection structure of the network with the
passivity properties of its components. Because the off-diagonal
components of this matrix are negative for links that represent
inhibitory reaction rates, diagonal stability is less restrictive
than a networked small-gain condition \cite{araki75,araki76} which
ignores the signs of the off-diagonal terms. In the case of a cyclic
graph where each link $l=1,\cdots,n$ connects source $i=l$ to sink
$i=l+1 \, ({\rm mod}\, n)$, and where only link $n$ has a negative
sign, (\ref{newE}) assumes the form (\ref{Ematrix}). Theorem
\ref{main} thus recovers the result of \cite{arcson06} as a special
case, and further relaxes it by accommodating the $g_i(x_i)$
functions in (\ref{newsys}) which are not allowed in
\cite{arcson06}.

\section{Examples}\label{examp}

\begin{example}\label{mapk}\rm
To illustrate Theorem \ref{main} we first study a simplified model
of mitogen activated protein kinase (MAPK) cascades with inhibitory
feedback, proposed in \cite{KholodenkoNegfeedback,stas2}:
\begin{eqnarray}
\dot{x}_1&=&-\frac{b_1x_1}{c_1+x_1}+\frac{d_1(1-x_1)}{e_1+(1-x_1)}\frac{\mu}{1+kx_3} \label{mapk1}\\
\dot{x}_2&=&-\frac{b_2x_2}{c_2+x_2}+\frac{d_2(1-x_2)}{e_2+(1-x_2)}x_1 \label{mapk2}\\
\dot{x}_3&=&-\frac{b_3x_3}{c_3+x_3}+\frac{d_3(1-x_3)}{e_3+(1-x_3)}x_2.
\label{mapk3}
\end{eqnarray}
The variables $x_i \in [0,1]$ denote the active forms of the
proteins, and the terms $1-x_i$ indicate the inactive forms (after
nondimensionalization and assuming that the total concentration of
each of the proteins is $1$). The second term in each equation
indicates the rate at which the inactive form of the protein is
being converted to active form, while the first term models the
inactivation of the respective protein. For the proteins $x_i$,
$i=2,3$, the activation rate is proportional to the concentration of
the active form of the protein $x_{i-1}$ upstream, which facilitates
the conversion. The activation of the first protein $x_1$, however,
is inhibited by $x_3$ as represented by the decreasing function
$\mu/(1+kx_3)$.

The model (\ref{mapk1})-(\ref{mapk3}) is of the form (\ref{newsys})
with
\begin{eqnarray}\nonumber
&&\!\!\!\!\!\!\!\!\!\! f_i(x_i)=\frac{b_ix_i}{c_i+x_i}, \
g_i(x_i)=\frac{d_i(1-x_i)}{e_i+(1-x_i)}, \ i=1,2,3, \\
&& \!\!\!\!\!\!\!\!\!\! h_i(x_i)=x_i, \ i=1,2, \quad
h_3(x_3)=\frac{\mu}{1+kx_3}.
\end{eqnarray}
Because the underlying graph is cyclic with each link $l=1,2,3$
connecting source $i=l$ to sink $i=l+1 ({\rm mod} 3)$, and because
$h_3(\cdot)$ is strictly decreasing, the dissipativity matrix $E$ in
(\ref{newE}) is of the form (\ref{Ematrix}) and, as proved in
\cite{arcson06}, its diagonal stability is equivalent to the secant
criterion (\ref{inc2c}). However, unlike the model (\ref{bio}) of
\cite{arcson06} which disallows state products, Theorem \ref{main}
above accommodates the functions $g_i(x_i)$, and is applicable to
(\ref{mapk1})-(\ref{mapk3}).

To reduce conservatism in the estimates for the $\gamma_i$'s in
Theorem \ref{main} we further restrict the intervals $[0,1]$ in
which $x_i$'s evolve by noting that $h_3(x_3)$ takes values within
the interval $[\frac{\mu}{1+k},\mu]$. Because $h_3(x_3)$ is the
input to the $x_1$-subsystem, and because the function
$\theta_i:[0,1]\rightarrow [0,\infty)$ defined by
\begin{equation}
\theta_i(x_i):=\frac{f_i(x_i)}{g_i(x_i)},
\end{equation}
is strictly increasing, it follows from the bounds on the input
signal that the interval
${\mathcal{X}}_1=[x_{1,min},x_{1,max}]:=[\theta_1^{-1}(\mu/(1+k)),\theta_1^{-1}(\mu)]$
is an invariant and attractive set for the $x_1$-subsystem. Since
$x_1$ and $x_2$  serve as inputs to the $x_2$- and $x_3$-subsystems
respectively, the same conclusion holds for the intervals
${\mathcal{X}}_2=[x_{2,min},x_{2,max}]$ and
${\mathcal{X}}_3=[x_{3,min},x_{3,max}]$, where
\begin{equation}
x_{i,min}:=\theta_{i}^{-1}(x_{i-1,min}) \quad
x_{i,max}:=\theta_{i}^{-1}(x_{i-1,max})
\end{equation}
$i=2,3$. With the following coefficients from \cite{Son02}:
\begin{eqnarray*}
&& \!\!\!\!\!\!\!\!\! b_1=e_1=c_1=b_2=0.1, \ c_2=e_2=c_3=e_3=0.01,
\\ && \!\!\!\!\!\!\!\!\! b_3=0.5,\ d_1=d_2=d_3=1,\ \mu=0.3,
\end{eqnarray*}
we obtained $\gamma_i$'s numerically by maximizing the left-hand
side of (\ref{growth}) on $\mathcal{X}_i$ for various values of the
parameter $k$. This numerical experiment showed that the secant
condition $\gamma_1\gamma_2\gamma_3<8$ is satisfied in the range
$k\le 4.35$ (for $k=4.36$ we get $\gamma_1\gamma_2\gamma_3=11.03$).
Reference \cite{Son02} gives a small-gain estimate $k\le 3.9$ for
stability, and shows that a Hopf bifurcation occurs at around
$k=5.1$. The estimate $k\le 4.35$ obtained from Theorem \ref{main}
thus reduces the gap between the unstable range and the small-gain
estimate.
\end{example}

\begin{example}\rm
A common form of feedback inhibition in metabolic networks occurs
when several end metabolites in different branches of a pathway
inhibit a reaction located before the branch point
\cite{stephanopoulous,ChiGroBas06}. As an example of this situation
we consider the network in Figure \ref{arb} where the end
metabolites with concentrations $x_4$ and $x_6$ inhibit the
formation of $x_1$ from an initial substrate $x_0$.
\begin{figure}[h]
\vspace{.8cm}
\begin{center}
\mbox{}\setlength{\unitlength}{0.8mm}
\begin{picture}(70,50)
\put(-10,-5){\psfig{figure=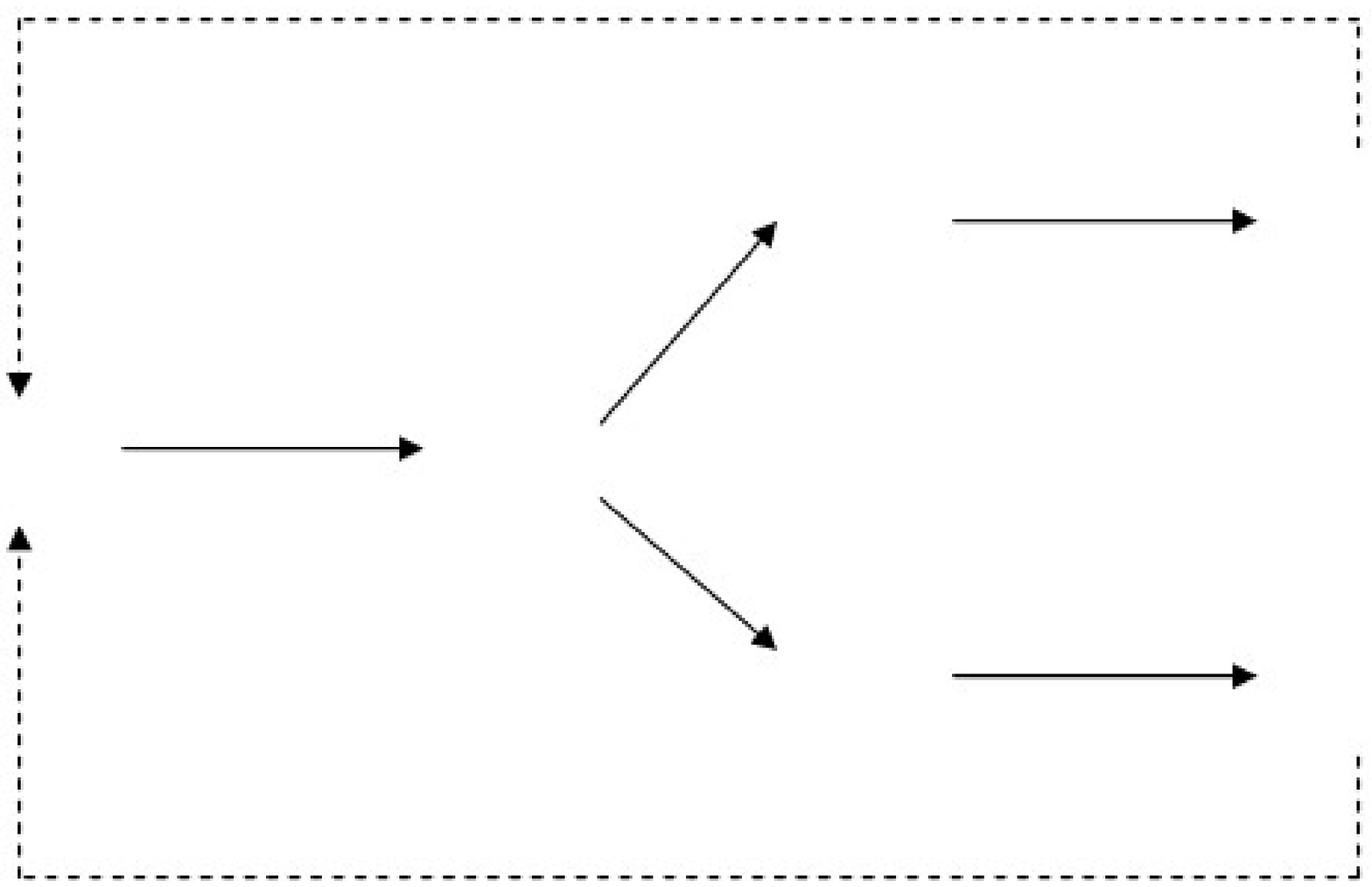,width=90\unitlength}}
 \put(-6,26){$x_1$}\put(22,26){$x_2$}\put(43.5,40){$x_3$}\put(72,40){$x_4$}\put(43.5,13){$x_5$}
 \put(72,13){$x_6$}\put(9,29.5){$1$}\put(31,35){$2$}\put(57,43.3){$3$}\put(35,55){$4$}\put(31,17){$5$}
 \put(57,8.5){$6$}\put(35,-3){$7$}
\end{picture}
 \caption{Feedback inhibition in a branched network. The dashed links $4$ and $7$ indicate negative (inhibitory)
 feedback signals. The dissipativity matrix obtained from (\ref{newE}) for this
 network is (\ref{dissipex}).
 } \label{arb}
\end{center}
\end{figure}
Assuming that $x_0$ is kept constant, and that its conversion to
$x_1$ is regulated by two isofunctional enzymes each of which is
selectively sensitive to $x_4$ or $x_6$, we represent this network
as in (\ref{newsys}):
\begin{eqnarray}
\nonumber
\dot{x}_1&=&-f_1(x_1)+h_4(x_4)+h_7(x_6)\\
\nonumber
\dot{x}_2&=&-f_2(x_2)+h_1(x_1)\\
\nonumber
\dot{x}_3&=&-f_3(x_3)+h_2(x_2)\\
\dot{x}_4&=&-f_4(x_4)+h_3(x_3)\\
\nonumber
\dot{x}_5&=&-f_5(x_5)+h_5(x_2)\\
\nonumber
 \dot{x}_6&=&-f_6(x_6)+h_6(x_5),
\end{eqnarray}
where the functions $h_4(x_4)$ and $h_7(x_6)$ are decreasing due to
the inhibitory effect of $x_4$ and $x_6$, while $h_l(\cdot)$,
$l=1,2,3,5,6$ and $f_i(\cdot)$, $i=1,\cdots,6$ are increasing.

Rather than study specific forms for these functions, we assume that
A1 and A2 hold, and that $\gamma_l$'s exist as in (\ref{growth}). An
application of Theorem \ref{main} then proves global asymptotic
stability  of the equilibrium if the dissipativity matrix
\begin{equation}\label{dissipex}
E=\left[\begin{array}{ccccccc}-\frac{1}{\gamma_1} & 0 & 0 & -1 & 0 &
0 & -1 \\ 1 & -\frac{1}{\gamma_2} & 0 & 0 & 0 & 0 & 0 \\ 0 & 1 &
-\frac{1}{\gamma_3} & 0 & 0 & 0 & 0 \\ 0 & 0 & 1 &
-\frac{1}{\gamma_4} & 0 & 0 & 0 \\ 1 & 0 & 0 & 0 &
-\frac{1}{\gamma_5} & 0 & 0 \\ 0 & 0 & 0 & 0 & 1 &
-\frac{1}{\gamma_6} & 0 \\ 0 & 0 & 0 & 0 & 0 & 1 &
-\frac{1}{\gamma_7}
\end{array} \right]
\end{equation}
is diagonally stable. Note that the $4\times 4$ principal
submatrices obtained by deleting row-column pairs
 $\{5,6,7\}$ and $\{2,3,4\}$ each exhibit a cyclic structure for
 which, as shown in \cite{arcson06}, diagonal stability is equivalent
 to the secant criteria
\begin{equation}\label{neces}
\gamma_1\gamma_2\gamma_3\gamma_4<\sec(\pi/4)^4=4 \quad \mbox{and}
\quad \gamma_1\gamma_5\gamma_6\gamma_7<4,
\end{equation}
respectively. Because principal submatrices of a diagonally stable
matrix are also diagonally stable, we conclude that (\ref{neces}) is
a necessary condition for the diagonal stability of
(\ref{dissipex}). In fact, we prove the following necessary and
sufficient condition:

\begin{lemma} The matrix $E$ in (\ref{dissipex}) is diagonally
stable iff
\begin{equation}\label{suft_par}
\gamma _1\gamma _2\gamma _3\gamma _4 + \gamma _1\gamma _5\gamma
_6\gamma _7 < \sec(\pi/4)^4 = 4 \,.
\end{equation}
\end{lemma}

\noindent \emph{Proof:} We prove the sufficiency of this condition
as a consequence of a more general fact. Consider the following
diagonal matrix:
\begin{equation}\label{diagD}
D = \mbox{diag}\,\left(
  1\,,\,\frac{\gamma _3\gamma _4}{2}\,,\,
        \frac{\gamma _4}{\gamma _2}\,,\,\frac{2}{\gamma _2\gamma _3}\,,\,
        \frac{\gamma _6\gamma _7}{2} \,,\,
        \frac{\gamma _7}{\gamma _5} \,,\, \frac{2}{\gamma _5\gamma _6}
  \right)\end{equation}
and the matrix $$M := E^TD+DE.$$ We will prove that
condition~(\ref{suft_par}) implies that $M\le 0$. Diagonal stability
of $E$ follows from this claim in view of the following argument:
Given any $\gamma_i$'s satisfying the constraint (\ref{suft_par}),
we can find  $\tilde{\gamma} _i>\gamma _i$ that still satisfy the
constraint, and under this transformation $E$ gets transformed to
$\tilde{E}=E+\Delta $, where $\Delta $ is some positive diagonal
matrix. Now let $\tilde{D}$ be defined for $\tilde{E}$ as in
(\ref{diagD}) with $\gamma_i$'s replaced by $\tilde{\gamma}_i$'s.
Since $E^T\tilde{D}+\tilde{D}E <
\tilde{E}^T\tilde{D}+\tilde{D}\tilde{E}=\tilde{M}$, and since
$\tilde{M}\le 0$, it follows that $E^T\tilde{D}+\tilde{D}E<0$, which
means that $E$ is diagonally stable.

To prove that (\ref{suft_par}) implies $M\le 0$, we let
$E_\varepsilon := E - \varepsilon I$  for each $\varepsilon >0$, and
show that $M_\varepsilon =E_\varepsilon ^TD+DE_\varepsilon $ is
negative definite for  small enough $\varepsilon >0$. By continuity,
this last property implies that $M\le 0$. In order to check negative
definiteness of $M_\varepsilon$, we consider the principal minors
$\mu _i(\varepsilon )$, $i=1,\ldots ,7$ of $M_\varepsilon $, and ask
that they all have sign $(-1)^i$ for small $\varepsilon >0$. Each
$\mu _i$ is a polynomial of degree $\leq 7$ on $\varepsilon$. The
determinant of $M_\varepsilon $ can be expanded as follows:
\begin{equation}
\mu _7(\varepsilon ) = \frac{8 \gamma _4 \gamma _7 (\gamma _5 + 2
\gamma _6 + \gamma _7) (\gamma _2 + 2 \gamma _3 + \gamma _4)}
   {\gamma _1 \gamma _2^3  \gamma _3 \gamma _5^3\gamma _6  } \,\Delta  \, \varepsilon ^2
 +\; O(\varepsilon ^3),
\end{equation}
where $ \Delta  = \gamma _1 \gamma _2 \gamma _3 \gamma _4 + \gamma
_1 \gamma _5 \gamma _6 \gamma _7 - 4 $. Similarly, we have:
\[
\mu _6(\varepsilon ) = \frac{-2 \gamma _4 \gamma _7^2  (\gamma _2 +
2 \gamma _3 + \gamma _4)}
  {\gamma _1 \gamma _2^3  \gamma _3 \gamma _5^2} \,\Delta
 \, \varepsilon
\;+\; O(\varepsilon ^2),
\]
\[
\mu _5(\varepsilon ) = \frac{2  \gamma _4 \gamma _6 \gamma _7(\gamma
_2 + 2 \gamma _3 + \gamma _4)}
  {\gamma _1 \gamma _2^3  \gamma _3 \gamma _5}  \,\Delta
 \, \varepsilon
\;+\; O(\varepsilon ^2),
\]
\[
\mu _4(\varepsilon ) = \frac{-2\gamma _4 (\gamma _2 + 2 \gamma _3 +
\gamma _4)}{\gamma _1 \gamma _2^3  \gamma _3} \,\Delta _1
 \, \varepsilon
\;+\; O(\varepsilon ^2),
\]
where $ \Delta _1 = \gamma _1 \gamma _2 \gamma _3 \gamma _4  - 4 $,
\[
\mu _3(\varepsilon ) = \frac{\gamma _4^2}{2 \gamma _1 \gamma _2^2}
\,\Delta _1 \;+\; O(\varepsilon ),
\]
\[
\mu _2(\varepsilon ) = \frac{-\gamma _3 \gamma _4 }{4 \gamma _1
\gamma _2} \, (\Delta _1-4) \;+\; O(\varepsilon ),
\]
and
\[
\mu _1(\varepsilon ) = -\frac{2}{\gamma _1}-2\varepsilon .
\]
Since $\Delta _1<\Delta $, we conclude that the matrix
$M_\varepsilon $ is negative definite for all small enough
$\varepsilon >0$ if and only if $\Delta <0$. In particular,
condition~(\ref{suft_par}) implies that $M\le 0$, as claimed.

Finally, we prove the necessity of (\ref{suft_par}) for the diagonal
stability of $E$ in (\ref{dissipex}). To this end, we define
$\hat{E}=\mbox{diag}\,\left(\gamma_1,\cdots,\gamma_7\right)E$ which
has all diagonal components equal to $-1$, and characteristic
polynomial equal to:
\[
(s+1)^3[(s+1)^4+k],
\]
where $k:=\gamma _1\gamma _2\gamma _3\gamma _4 + \gamma _1\gamma
_5\gamma _6\gamma _7$. For $k\ge0$, the roots of $(s+1)^4=-k$ have
real part
   $\pm \sqrt[4]{k/4} - 1$; hence $k<4$ is necessary for these real parts to be
   negative.
Because (\ref{suft_par}) is necessary for the Hurwitz property of
$\hat{E}$, it is also necessary for its diagonal stability. Since
diagonal stability of $\hat{E}$ is equivalent to diagonal stability
of $E$, we conclude that (\ref{suft_par}) is necessary for the
diagonal stability of ${E}$.
\end{example}

\section{Stability of a Compartmental Model with
Diffusion}\label{compsec} A compartmental model is appropriate for
describing the spatial localization of processes when each of a
finite set of spatial domains (``compartments") is well-mixed, and
can be described by ordinary differential equations. Instead of the
lumped model (\ref{newsys}), we now consider $n$ compartments, and
represent their interconnection structure with a new graph in which
the links $k=1,\cdots,m$ indicate the presence of diffusion between
the compartments $j=1,\cdots,n$ they interconnect. Although the
graph is undirected, for notational convenience we assign an
orientation to each link and define the $n\times m$ \emph{incidence
matrix} $S$
 as \setlength{\arraycolsep}{0.15em}
\begin{equation}s_{jk}:=\left\{
\begin{array}{cl}
+1&\mbox{if node $j$ is the sink of link $k$}\\
-1&\mbox{if node $j$ is the source of link $k$}\\
0&\mbox{otherwise}.\label{dik}\\
\end{array}
\right.
\end{equation} The particular
choice of the orientation does not change the derivations below.

We first prove a general stability result (Theorem \ref{compar}
below) for a class of compartmental models interconnected as
described by the incidence matrix $S$. We then apply this result in
Corollary \ref{coro} to the situation where the individual
compartments possess dynamics of the form
 studied in Section \ref{mainsec}.
We let
$$X_j:=(x_{j,1},\cdots,x_{j,N})^T$$
be the state vector of concentrations $x_{j,i}$ in compartment $j$,
and let $\dot{X}_j=F_j(X_j)$ represent the dynamics of the $j$th
compartment in the absence of diffusion terms. Next, for each link
$k=1,\cdots,m,$ we denote by
\begin{equation}\label{diffterms} \mu_{k,i}(x_{{\rm sink}(k),i}-x_{{\rm source}(k),i})
\end{equation}
the diffusion term for the species $i$, flowing from ${\rm
source}(k)$ to ${\rm sink}(k)$, and assume the functions
$\mu_{k,i}(\cdot)$, $k=1,\cdots,m$, $i=1,\cdots,N$, satisfy
\begin{equation}
    \sigma \mu_{k,i}(\sigma)
    \, \leq \, 0,
    ~~
    \forall \, \sigma \, \in \, \bbR.
    \label{eq.C7}
    \end{equation}
Then, the coupled dynamics of the compartments become:
\begin{equation}
    \label{eq.CM}
\dot{X}_j=F_j(X_j)+(S_{j,\cdot}\otimes I_N)\mu((S^T\otimes I_N)X)
\quad j=1,\cdots,n
\end{equation}
where $S_{j,\cdot}$ is the $j$th row of the incidence matrix $S$,
$I_N$ is the $N\times N$ identity matrix, ``$\otimes$" represents
the Kronecker product,
\begin{equation}\label{concat}
X:=[X_1^T \cdots X_n^T]^T
\end{equation}
and $\mu:\bbR^{mN}\rightarrow \bbR^{mN}$ is defined as
\begin{equation}\label{diagop}
\mu(z):= [\mu_{1,1}(z_1) \cdots \mu_{1,N}(z_N)\ \cdots \ \cdots \
\mu_{m,1}(z_{(m-1)N+1})\cdots \mu_{m,N}(z_{mN})]^T_.
\end{equation}
We now prove stability of the coupled system (\ref{eq.CM}) under the
assumption that a common Lyapunov function exists for the decoupled
models $\dot{X}_j=F_j(X_j)$, $j=1,\cdots,n$, and that this common
Lyapunov function consists of a sum of convex functions of
individual state variables:

\begin{theorem}\rm \label{compar}
Consider the system (\ref{eq.CM}) where the function $\mu(\cdot)$ is
as in (\ref{diagop}) and (\ref{eq.C7}). If there exists a Lyapunov
function $V:\bbR^N\rightarrow \bbR$ of the form
\begin{equation}\label{deco}
V(x)=V_1(x_1)+\cdots+V_N(x_N)
\end{equation}
where each $V_i(x_i)$ is a convex, differentiable and positive
definite function, satisfying
\begin{equation}\label{uncoupled}
\nabla V(x) F_j(x) \le -\alpha(|x|) \quad j=1,\cdots,n
\end{equation}
for some class-$\mathcal{K}$ function $\alpha(\cdot)$, then the
origin $X=0$ of (\ref{eq.CM}) is asymptotically stable. If, further,
$V(\cdot)$ is radially unbounded, then $X=0$ is globally
asymptotically stable.

\end{theorem}

\bigskip

\noindent \emph{Proof:} We employ the composite Lyapunov function
\begin{equation}
\mathcal{V}(X) \; = \; \sum_{j=1}^{n}V(X_j),
\end{equation}
and obtain from (\ref{eq.CM}) and (\ref{uncoupled}):
\begin{equation}
\dot{\mathcal{V}}(X)\le -\sum_{j=1}^n \alpha(|X_j|)+ [\nabla
V(X_1)\cdots \nabla V(X_n)](S\otimes I_N)\mu((S^T\otimes
    I_N)X).
    \label{derv1}
\end{equation}
We next rewrite the second term in the right-hand side of
(\ref{derv1}) as
\begin{equation}\label{rewtrans}
\left((S^T\otimes I_N)\left[\begin{array}{c}\nabla V^T(X_1)\\ \vdots \\
\nabla V^T(X_n)\end{array}\right]\right)^T\mu((S^T\otimes
    I_N)X),
\end{equation}
and note from (\ref{dik}) that (\ref{rewtrans}) equals
\begin{equation}\label{rew2}
\sum_{k=1}^m [\nabla V^T(X_{{\rm sink}(k)})-\nabla V^T(X_{{\rm
source}(k)})]\left[\begin{array}{c} \mu_{k,1}\\ \vdots
\\ \mu_{k,N}\end{array} \right]
\end{equation}
where $\mu_{k,i}$, $i=1,\cdots,N$, denotes the diffusion function
(\ref{diffterms}), and the argument  is dropped for brevity. Next,
using (\ref{deco}), we rewrite (\ref{rew2}) as
\begin{equation}\label{re2}
\sum_{k=1}^{m}\sum_{i=1}^N [\nabla V_i(x_{{\rm sink}(k),i})-\nabla
V_i(x_{{\rm source}(k),i})]\,\mu_{k,i}.
\end{equation}
Because $V_i(\cdot)$ is a convex function, its derivative $\nabla
V_i(\cdot)$ is a nondecreasing function and, hence, $\nabla
V_i(x_{{\rm sink}(k),i})-\nabla V_i(x_{{\rm source}(k),i})$
possesses the same sign as $(x_{{\rm sink}(k),i}-x_{{\rm
source}(k),i})$. We next recall from the sector property
(\ref{eq.C7}) that the function $\mu_{k,i}$ in (\ref{diffterms})
possesses the opposite sign of its argument $(x_{{\rm
sink}(k),i}-x_{{\rm source}(k),i})$. This means that each term in
the sum (\ref{re2}) is nonpositive
 and, hence, (\ref{derv1}) becomes
\begin{equation}\label{re3}
 \dot{\mathcal{V}}(x)\le -\sum_{j=1}^n \alpha(|X_j|),
\end{equation}
from which the conclusions of the theorem follow. \hfill $\Box$

\medskip

Theorem \ref{compar} is applicable when each compartment is as
described in Section \ref{mainsec}, $h_l(\cdot)$ satisfies
(\ref{recp}), and $g_i(\cdot)$'s, $i=1,\cdots,N$, are nonincreasing
functions. This is because the Lyapunov construction (\ref{lyap}) in
Section \ref{mainsec} consists of a sum of terms as in (\ref{deco}),
each of which is convex when the derivative of (\ref{comp}) is
nondecreasing:

\begin{corollary}\rm \label{coro}
Consider the system (\ref{eq.CM}) where the function $\mu(\cdot)$ is
as in (\ref{diagop}) and (\ref{eq.C7}), and $F_j(x)$,
$j=1,\cdots,n$, are identical and represent the right-hand side of
(\ref{newsys}). If all assumption of Theorem \ref{main} hold and if,
in addition, $h_l(\cdot)$ satisfies (\ref{recp}), and
$g_i(\cdot)$'s, $i=1,\cdots,N$, are nonincreasing functions, then
the equilibrium $X=[x^{*T},\cdots,x^{*T}]^T$ is globally
asymptotically stable.
\end{corollary}

\section{Comparison of the State-Space and Input/Output Approaches}
\label{MoyHillVid} The earlier paper~\cite{secant} gave a purely
input/output (instead of state-space) version of the secant
criterion, phrased in the language of passivity of $L^2$ operators.
We now explain how to extend this I/O approach to the general graphs
studied in this paper. The result follows easily by imposing an
appropriate diagonal stability condition, combined with a  key lemma
due to
 Moylan and Hill~\cite{MoyHill78}, and  Vidyasagar \cite{vidyasagar2}.
Below we give a streamlined version of this lemma, and compare it
with the state space approach employed earlier in this paper.

We denote by $L^2_e$ the extended space of signals (thought of as
time functions) $w:[0,\infty )\rightarrow \R$ which have the
property that each restriction  $w_T = w|_{[0,T]}$ is in $L^2(0,T)$,
for every $T>0$. Given an element $w\in L^2_e$ and any fixed $T>0$,
we write $\normT{w}$ for the $L^2$ the norm of this restriction
$w_T$, and given two functions $v,w\in L^2_e$ and any fixed $T>0$,
the inner product of $v_T$ and $w_T$ is denoted by $\ipT{v}{w}$. The
same notation is  used for vector functions.

We view the $M$ subsystems to be interconnected as operators $\Sigma
_i: L^2_e\rightarrow L^2_e : u_i\mapsto y_i$, and impose the
following strict passivity property: there exist constants $\gamma
_i>0$ (``secant gains'' in~\cite{secant}) such that
\begin{equation}\label{uTyT}
\normT{y_i}^2 \;\leq \; \gamma _i\ipT{y_i}{u_i} \;\;\mbox{for each}
\; i=1,\ldots ,M \;\;\mbox{and each}  \; T>0\,.
\end{equation}
We then consider the interconnection where
\begin{equation}\label{uvy}
u_i(t) = v_i(t) + A_iy(t)\,,
\end{equation}
or just $u=v+Av$, where the $v_i$'s are external inputs,
$y=\mbox{col}(y_1,\ldots ,y_M)$, $v=\mbox{col}(v_1,\ldots ,v_M)$,
and the $A_i$, $i=1,\ldots ,M$ are the rows of an interconnection
matrix $A\in \R^{M\times M}$.  In other words, the $i$th subsystem
receives as inputs an external input plus an appropriate linear
combination of outputs from the remaining systems (including
possibly feedback from itself, if the corresponding diagonal entry
of $A$ is nonzero). We introduce:
\[
E\,:=\; A \,-\, \Gamma
\]
where $\Gamma  = \mbox{diag}\,(\frac{1}{\gamma _1},\ldots
,\frac{1}{\gamma _M})$.

\noindent{\bf Lemma.} Suppose that there exists a diagonal positive
definite matrix $D\in \R^{M\times M}$ such that
\[
DE + E'D < 0\,.
\]
Then, the system obtained from the systems $\Sigma _i$ using the
interconnection matrix $A$ is $L^2$ stable as a system with input
$v$ and output $y$. More precisely, there is some constant $\rho >0$
such that, for any $u,v,y\in (L^2_e)^M$ such that~(\ref{uTyT}) and
~(\ref{uvy}) hold, necessarily $\normT{y}\leq \rho \normT{v}$ for
all $T>0$ (and therefore also $\norm{y}\leq \rho \norm{v}$, if $v\in
(L^2)^M$).

\noindent \emph{Proof:} We pick an $\alpha >0$ such that $DE + E'D <
-2\alpha I$, and observe that, for any $T>0$ and any function $z\in
L^2(0,T)$, it holds that
\[
\ip{Dz}{Ez} \,=\, \int_0^T z(s)'DEz(s)\,ds \,=\, \int_0^T
\frac{1}{2} z'(s)(DE + E'D)z(s)\,ds \,\leq \, -\alpha \int_0^T
z'(s)z(s)\,ds \,=\,  -\alpha \norm{z}^2.
\]
Fix an arbitrary $T>0$, and write $D=\mbox{diag}\,(d_1,\ldots
,d_M)$. Since, for each $i$, $\ipT{y_i}{u_i - \frac{1}{\gamma
_i}y_i}\geq 0$, it follows that also $\ipT{d_iy_i}{u_i -
\frac{1}{\gamma _i}y_i}\geq 0$, or, in vector form:
\[
\ipT{Dy}{u - \Gamma y}\geq 0.
\]
Substituting $u=v+Ay$, we obtain: $\ipT{Dy}{v + Ey}\geq 0$, from
which, using the Cauchy-Schwartz inequality:
\[
\beta \normT{v}\normT{y} \,\geq \, \ipT{Dy}{v} \,\geq \,
-\ipT{Dy}{Ey} \,\geq \, \alpha \normT{y}^2
\]
for some $\beta >0$. So $\normT{y}\leq \rho \normT{u}$, with $\rho
=\frac{\beta}{\alpha}$, as desired. \hfill $\Box$

\medskip

State-space stability results may be obtained as corollaries, by
combining this I/O result with appropriate detectability and
controllability conditions, as discussed in~\cite{secant}.  However,
the direct Lyapunov approach employed earlier in this paper allowed
us to formulate verifiable state-space conditions that guarantee the
desired passivity properties for the subsystems. These conditions
are particularly suitable for systems of biological interest because
they are applicable to models with nonnegative state variables, and
do not rely on the knowledge of the location of the equilibrium. The
state-space approach further made it possible to prove robustness of
our stability criterion in the presence of diffusion terms.

\section{Conclusions} \label{conc}
We have presented a passivity-based stability criterion for a class
of interconnected systems, which encompasses the secant criterion
for cyclic systems \cite{arcson06} as a special case. Unlike the
result in \cite{arcson06}, we have further allowed the presence of
state products in our model.  Our main result (Theorem \ref{main})
determines global asymptotic stability of the network from the
diagonal stability of the dissipativity matrix (\ref{newE}) which
incorporates information about the passivity properties of the
subsystems, the interconnection structure of the network, and the
signs of the interconnection terms. Although diagonal stability can
be checked numerically with efficient linear matrix inequality (LMI)
tools \cite{boyd},  it is of interest to derive analytical
conditions that make explicit the role of the reaction rate
coefficients on stability properties. Indeed our earlier paper
\cite{arcson06} showed that the diagonal stability of negative
feedback cyclic systems is equivalent to the secant criterion of
\cite{tysoth78,thr91}. In Example 2 we have derived a similar
analytical condition for a branched cyclic interconnection
structure. Further studies for deriving analytical conditions for
practically important interconnection structures would be of great
interest. Another research topic is to extend the stability result
for compartmental models with diffusion in Section \ref{compsec} to
partial differential equation models. On this topic we have reported
preliminary results applicable to cyclic systems in
\cite{07acc_jovanovic_arcak_sontag}, and are currently studying more
general interconnection structures.

\bibliographystyle{unsrt}       

\end{document}